\documentclass[twocolumn,english,aps,prb,floatfix,preprintnumbers,showpacs,amsfonts,amssymb,superscriptaddress]{revtex4}
\usepackage{ae,aecompl}
\usepackage[T1]{fontenc}
\usepackage[latin1]{inputenc}
\usepackage{amsmath}
\usepackage{graphicx}
\usepackage{amssymb}

\makeatletter

\providecommand{\tabularnewline}{\\}

\@ifundefined{textcolor}{}
{%
 \definecolor{BLACK}{gray}{0}
 \definecolor{WHITE}{gray}{1}
 \definecolor{RED}{rgb}{1,0,0}
 \definecolor{GREEN}{rgb}{0,1,0}
 \definecolor{BLUE}{rgb}{0,0,1}
 \definecolor{CYAN}{cmyk}{1,0,0,0}
 \definecolor{MAGENTA}{cmyk}{0,1,0,0}
 \definecolor{YELLOW}{cmyk}{0,0,1,0}
 }

\usepackage{amscd}
\usepackage{bm}
\usepackage{graphics,psfrag}
\usepackage{graphicx,psfrag}

\makeatother

\usepackage{babel}

\begin{document}

\title{Precise Determination of Quantum Critical Points by the Violation
of the Entropic Area Law}

\author{J.~C.~Xavier}

\affiliation{Instituto de F\'{\i}sica, Universidade Federal de Uberlândia, Caixa
Postal 593, 38400-902 Uberlândia, MG, Brazil }

\affiliation{Instituto de F\'{\i}sica de São Carlos, Universidade de São Paulo,
Caixa Postal 369, 13560-970 São Carlos, SP, Brazil}

\author{F.~C.~Alcaraz}

\affiliation{Instituto de F\'{\i}sica de São Carlos, Universidade de São Paulo,
Caixa Postal 369, 13560-970 São Carlos, SP, Brazil}

\date{\today{}}
\begin{abstract}
Finite-size scaling analysis turns out to be a powerful tool to calculate
the phase diagram as well as the critical properties of two dimensional
classical statistical mechanics models and quantum Hamiltonians in
one dimension. The most used method to locate quantum critical points
is the so called \textit{crossing method}, where the estimates are
obtained by comparing the mass gaps of two distinct lattice sizes.
The success of this method is due to its simplicity and the ability
to provide accurate results even considering relatively small lattice
sizes. In this paper, we introduce an estimator that locates quantum
critical points by exploring the known distinct behavior of the entanglement
entropy in critical and non critical systems. As a benchmark test,
we use this new estimator to locate the critical point of the quantum
Ising chain and the critical line of the spin-1 Blume-Capel quantum
chain. The tricritical point of this last model is also obtained.
Comparison with the standard crossing method is also presented. The
method we propose is simple to implement in practice, particularly
in density matrix renormalization group calculations, and provides
us, like the crossing method, amazingly accurate results for quite
small lattice sizes. Our applications show that the proposed method
has several advantages, as compared with the standard crossing method,
and we believe it will become popular in future numerical studies.
\end{abstract}


\maketitle

\section{Introduction}

In recent years, the connection between the quantum correlations and
the entanglement properties of quantum critical systems motivated
the research on the possible characterization of critical phenomena
via quantum information concepts. \citep{naturefazio,praosborne,cvidal,revfazio,prlkorepin,cardyentan,chicosarandy}
In particular, several of these studies have been devoted to the identification
of possible measures of entanglement that give a precise localization
of the quantum critical point in quantum chains. These studies basically
considered local measures of entanglement like the concurrence and
the negativity,\citealp{chicosarandy,PhysRevA.80.010304,PhysRevA.69.022107,PhysRevA.69.054101}
or other measures of quantum correlations that are not always simple
to calculate.\citealp{PhysRevLett.88.017901,PhysRevLett.92.027901,PhysRevLett.97.170401,PhysRevB.78.224413,PhysRevA.80.022108,PhysRevLett.105.095702}
The concurrence and the negativity are easy to measure, however can
not be considered universal measures, since depending on the distance
of the sites they are measured, and the model, they may detect or
not the critical points.\citealp{naturefazio,praosborne,PhysRevA.69.022107,PhysRevA.81.032334}

On the other hand, the standard theory of finite-size scaling (FSS)
gives us a quite simple way to locate quantum phase transitions. The
procedure, usually named as the \textit{crossing method} (CM), is
based on the crossing of the functions $LG(L)=L'G(L')$ at the critical
points,\citealp{barber} where $G(L)$ and $G(L')$ are the mass gaps
of the quantum chains with lattice sizes $L$ and $L'$, respectively.
This is the most used method to locate a phase transition in a quantum
chain. The success of this method is due to the fact that it provides
precise estimates of critical points even considering \emph{small
system sizes}. This point is very important because, with the exception
of the exactly soluble models, we do not have access of the ground
state properties of large systems. Normally, for a given lattice size,
the CM method gives the most accurate localization of critical points.

A natural question concerns the existence of an universal estimator,
based on the changing of entanglement at the quantum critical points,
that gives the same order of precision, for a given numerical effort,
as the standard CM. In this paper, we are going to present such estimator
for the quantum chains. The estimator will be obtained by exploiting
the well known violation of the area law of the entanglement entropy
of a subsystem with the rest of the quantum chain that happens in
a critical quantum point.

The entropy itself could, in principle, be used to locate the quantum
critical points. However, as we are going to show in two examples,
we need to use much larger lattices, to achieve the same precision,
as compared with the CM. The estimator we propose is given by the
difference of the entanglement entropy when we split the quantum chain
in subsystems of different sizes. As a benchmark test, we are going
to use this estimator to determine the localization of critical points
of two well known models: the quantum Ising chain and the spin-1 Blume-Capel
model (BCM). These are good examples since in the Ising model we have
a well known critical point, and in the BCM we possess a line of second
order phase transition in the Ising universality, that ends in a tricritical
point in the universality class of the tricritical Ising model. Surprisingly,
the estimates for the critical and tricritical points obtained by
using the proposed estimator are as precise as the standard CM, for
the same lattice sizes. Moreover, the new method in comparison to
the CM has some advantages. We need only to calculate the ground state
eigenfunction, instead of the mass gaps as in the CM. Furthermore
at the same time as the quantum critical point is located the conformal
anomaly of the underlying conformal field theory governing the critical
point is also evaluated. Our applications also shows that in the case
of first order transitions (gapped) the proposed estimator, differently
from the CM, shows a distinct finite-size behavior already for quite
small lattices, that in principle can be used to detect if the phase
transition is discontinuous (first order) or not.

The paper is organized as follows. Since the proposed method to locate
quantum phase transitions is going to be compared with the standard
CM, we present both methods in the next section. In the sections III
and IV, we present our results for the quantum Ising chain and the
BCM, respectively. Finally, in section V we summarize our results
and present our conclusions.

\section{Estimators of quantum critical points}

Let us consider a quantum chain whose Hamiltonian $H(\lambda)$ has
a quantum critical point at $\lambda=\lambda_{c}$. The FSS theory
tell us that at the quantum critical point, the finite chain with
$L$ sites will have a mass gap $G_{L}(\lambda)\sim1/L$, as $L\to\infty$.
This gives us the most popular method used to locate phase transitions.
The finite-size estimates of the critical point ($\lambda_{L,L'}$)
are obtained from the mass gaps of \emph{two distinct lattice sizes}.
They are obtained from the crossing of the function $F_{L}(\lambda)=LG_{L}(\lambda)$
for two lattice sizes,\citep{commgap} i. e.,

\begin{equation}
LG_{L}(\lambda)|_{\lambda_{L,L'}}=L'G_{L'}(\lambda)|_{\lambda_{L,L'}},\label{eq:1}\end{equation}
For this reason, this method is also known as the CM. Its application
is quite simple because it is necessary to evaluate only the two lowest
eigenvalues of $H(\lambda)$. A difficulty of this method happens
in the case where the quantum chain undergoes a first-order phase
transition with a small (but finite) mass gap $G_{\infty}$. In this
case, we also find crossing points $\lambda_{L,L'}$, since for $L<1/G_{\infty}$
we should expect $G_{L}(\lambda)\sim1/L$, as in a true critical point.

The purpose in this paper is to introduce and test a alternative method
to locate critical points in quantum chains. The method we propose
exploits the changing of behavior of the entanglement entropy of a
segment of the quantum chain, as we cross a quantum critical point.

In order to present our method, let us first define the well known
entanglement entropy. Consider a quantum chain with $L$ sites, described
by a pure state whose density operator is $\rho$. Let us consider
that the system is composed by the subsystems ${\cal A}$ with $\ell$
sites ($\ell=1,\ldots,L$) and ${\cal B}$ with $L-\ell$ sites. The
entanglement entropy is defined as the von Neumann entropy $S_{L}(\ell)=-\mbox{Tr}\rho_{{\cal {A}}}\ln\rho_{{\cal {A}}}$,
associated to the reduced density matrix $\rho_{{\cal {A}}}=\mbox{Tr}_{{\cal {B}}}\rho$. 

Most of the critical chains besides being scale invariant are also
conformal invariant, its long-distance physics is governed by a conformal
field theory (CFT) with central charge $c$. In the scaling regime
$1<<\ell<<L$ the entanglement entropy $S_{L}(\ell,\lambda)$ of the
ground state behaves differently if the system is critical ($\lambda=\lambda_{c}$)
or not. It increases logarithmically with the correlation length $\zeta_{L}(\lambda)$,
or equivalently, with the decrease of the mass gap $G_{L}(\lambda)=1/\zeta_{L}(\lambda)$,
namely:\citealp{cold,cardyentan,entroreviewcalabrese,cvidal,affleckboundary}

\begin{equation}
S_{L}(\ell,\lambda)=\begin{cases}
\gamma\ln\left[\frac{L}{\pi}\sin\left(\frac{\pi\ell}{L}\right)\right]+\beta, & \lambda=\lambda_{c}\\
-\gamma\ln\left[G_{L}(\lambda)\right]+\beta', & \lambda\ne\lambda_{c}\end{cases},\label{eq:entropy}\end{equation}
where $\gamma$ is an universal constant related with the central
charge $c$, namely $\gamma=c/3$ for periodic chains and $\gamma=c/6$
for chains with open boundaries. The constants $\beta$ and $\beta'$
are non universal and model dependent. Here, we consider only PBC in 
order to avoid boundary effects. 

The estimator we propose to locate a quantum critical point $\lambda_{c}$
is given by the coupling $\lambda_{c}^{L}$ that gives the maximum
value of the entanglement entropy difference (MVEED) given by:

\begin{equation}
\Delta S_{L}(\lambda)=S_{L}(\ell,\lambda)-S_{L}(\ell',\lambda),\label{eq:diff}\end{equation}
with $\ell=L/2$ and $\ell'=L/4$.
According to Eq. (\ref{eq:entropy}), as $L\to\infty$, we should
have 

\begin{equation}
\Delta S_{L}(\lambda)=\begin{cases}
\frac{\gamma}{2}\ln(2), & \lambda=\lambda_{c}\\
0, & \lambda\ne\lambda_{c}\end{cases}.\label{eq:diff2}\end{equation}

This means that as $L$ increases the estimator goes to zero, except
at the critical point, where its value is proportional to the central
charge $c$. Although in the definition (\ref{eq:diff}) it is assumed
that the lattice sizes are multiples of 4, for other lattice sizes
a simple variation of Eq. (\ref{eq:diff}) can also be defined.
In principle, we could choose two arbitrary subsystem sizes $\ell$ and $\ell'$. 
However, the signature of the violation of  the entropic area law is
observed more clear for $\ell,\ell'>>1$. Then, we have to choose 
$\ell,\ell'\sim L$. On the other hand, we also want to distinguish 
$\Delta S_L(\ell,\ell')$ from zero. 
This means that we have to choose small values of $\ell'$, when compared
with $\ell\sim L$, and satisfying $\ell'>>1$. So, it seems to be natural 
to choose $\ell=L/2$ and $\ell'=L/4$.

In the following sections, we are going to use the MVEED estimator
to locate the critical points of two different models, and compare
them with those obtained through the standard CM. Note that in the
same line, we could, in principle, also use as estimators the difference
of the $\alpha$-Rényi entropies. However, for $\alpha>1$ unusual
corrections appears\citep{entropyosc} and the finite-size effects
became more relevant (see also Refs. \onlinecite{xavieralcarazosc,taddiaosc}).

The key ingredient in the MVEED method is the violation of the entropic
area law at the critical point. This violation have been observed
in several systems such as the conformal invariant critical chains.
We could naively expect that the entanglement entropy of a $d$-dimensional
system would behave as $S\sim L^{d}$, since the entropy is usually
an extensive quantity. However this is not correct, the entanglement
entropy of the subsystems ${\cal A}$ and $B$, with arbitrary volumes,
are identical, i. e.: $S_{A}=\mbox{-Tr}\rho_{{\cal {A}}}\ln\rho_{{\cal {A}}}=S_{B}=-\mbox{Tr}\rho_{{\cal {B}}}\ln\rho_{{\cal {B}}}$.
Since both subsystems share the same area, we should expect the area-law
dependence of the entropy, i. e., $S_{A}=S_{B}\sim L^{d-1}$.\citep{PhysRevLett.71.666,PhysRevA.73.012309,PhysRevLett.94.060503}
This means that the information is shared only among the degrees of
freedom localized around the surface separating both systems. However,
at the critical point the correlation length diverges and both subsystems
share much more information among themselves. We thus expect, in this
case, a violation of the entropic area law. As we already mention,
the one dimension conformal invariant critical systems {[}see Eq.
(2){]} are celebrated examples where this law is violated. There are
also examples, in dimension $d>$1, where the entropic area law is
violated, such as some gapless fermionic systems with  a finite Fermi
surface. \citep{PhysRevLett.96.010404,PhysRevLett.96.100503,PhysRevLett.100.215701,PhysRevA.74.022329,PhysRevB.74.073103,PhysRevLett.105.050502,farkas}
It is interesting also to point out that in some disordered one-dimensional
systems the entropic law can be violated in a different way than the
one presented in Eq. (\ref{eq:entropy}).\citealp{1367-2630-12-11-113049}

We should mention that the difference of the entanglement entropy
of chains with \emph{distinct sizes} were used to extract the central
charge $c$ in one dimension quantum systems with good accuracy. \citep{PhysRevB.73.024417,xavierentanglement,PhysRevB.83.195105,rencentralcharge}
Läuchli and Kollath in Ref. \onlinecite{Kollath} also used a similar
procedure to detect the continuum line of critical points of the bosonic
Hubbard chain. In their procedure they considered the following estimator:
$\Delta S_{LK}(L,L'=L/2,\lambda)=S_{L}(L/2,\lambda)-S_{L/2}(L/4,\lambda)$.

For the above model, and also for the $XXZ$ quantum chain,
where a continuum line of critical points appears
by tuning a single coupling, the difference of entropies tends toward
a constant, appearing a plateau along the critical line. In the non
critical region they go to zero, as the lattice size increases. In
this case, we should not use the maximum of $\Delta S_{L}(\lambda)$ {[}or
the maximum of $\Delta S_{LK}(\lambda)${]} but its over all size
dependence to locate the endpoint of the critical line. We also have
calculated  $\Delta S_{L}(\lambda)$ for the $XXZ$ chain, for $L<100$, and
we confirmed that the end point of the critical line is better estimated 
by the overall sizes dependence of $\Delta S_{L}(\lambda)$, as compared with
the MVEED.


For the sake of comparison, we also test the estimator proposed by Läuchli and
Kollath to detect the critical point of the quantum Ising chain. As
our estimator, $\Delta S_{LK}(L,L',\lambda)$ also has a maximum close
to the critical point {[}see next section and Fig. 1(e){]}. So, it
is also possible to estimate the critical point by considering the
coupling that gives the maximum values of $\Delta S_{LK}(L,L')$.
However, note that to obtain one estimate of the critical point, with this
estimator, we need to consider two distinct lattice sizes, one being
twice the size of the other. 
This demand larger numerical effort. Besides that we verify, for both
models with $L=16$, that the errors of our estimates are approximately
two orders of magnitude smaller than the ones obtained by
considering the estimator $\Delta S_{LK}(L,L'=L/2,\lambda)$. \cite{commx}

\section{Quantum Ising chain - critical point determination}

The Ising quantum chain describes the dynamics of spin-1/2 localized
spins whose Hamiltonian is given by

\begin{equation}
H_{Ising}=-\sum_{j}\left(\sigma_{j}^{x}\sigma_{j+1}^{x}+\lambda\sigma_{j}^{z}\right)\label{eq:ising}\end{equation}
where $\sigma^{x}$, $\sigma^{z}$ are spin-1/2 Pauli matrices. It
depends on the parameter $\lambda$, and for simplicity, hereafter
we are going to consider only periodic chains. This model has a critical
point $\lambda_{c}=1$ that can be obtained from its exact solution,
or even more simply from its self dual property.\citep{Kogut1}

\begin{figure}
\begin{centering}
\psfrag{ell}{$\ell$}
\includegraphics[scale=0.18]{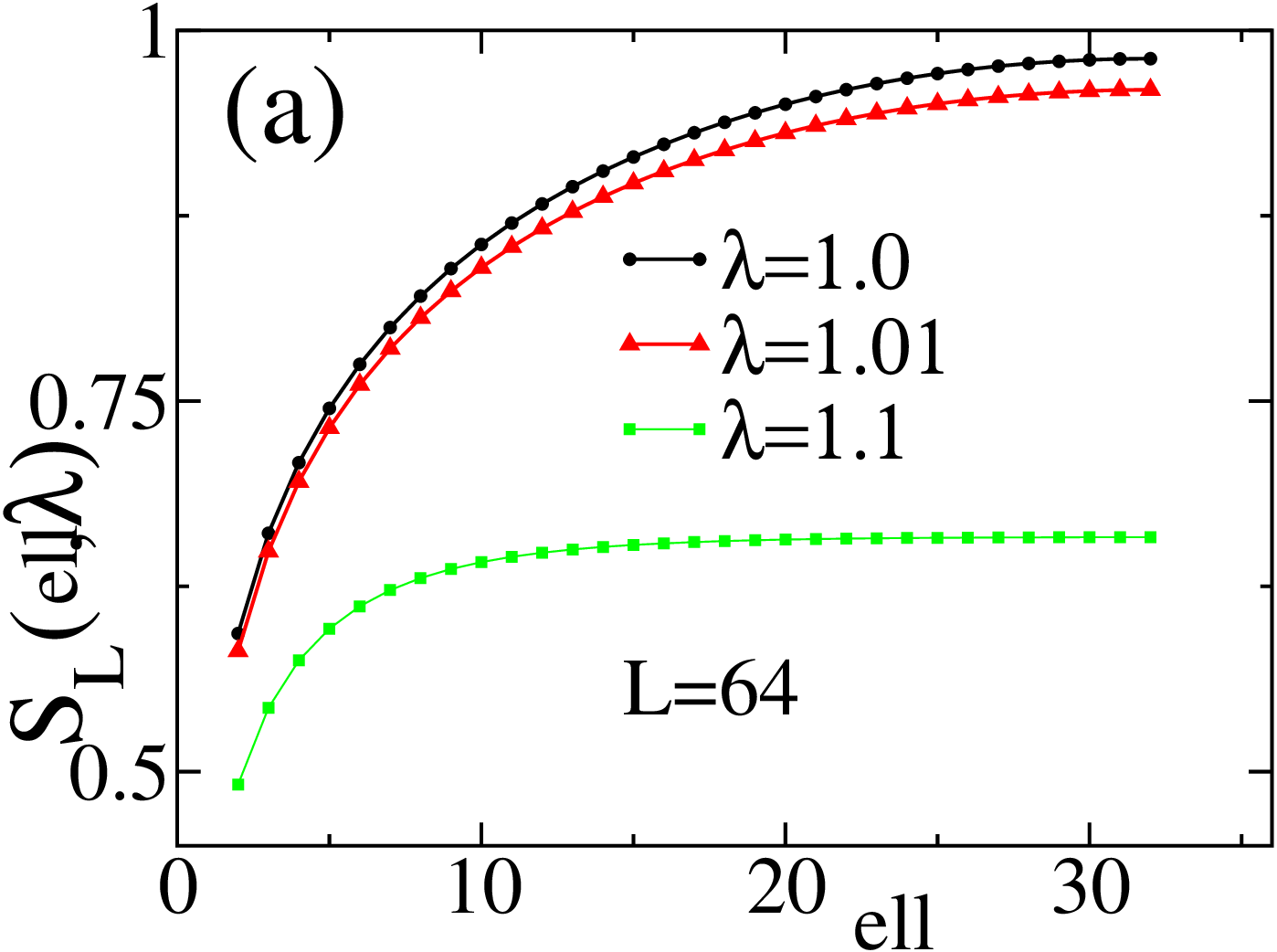}\includegraphics[scale=0.18]{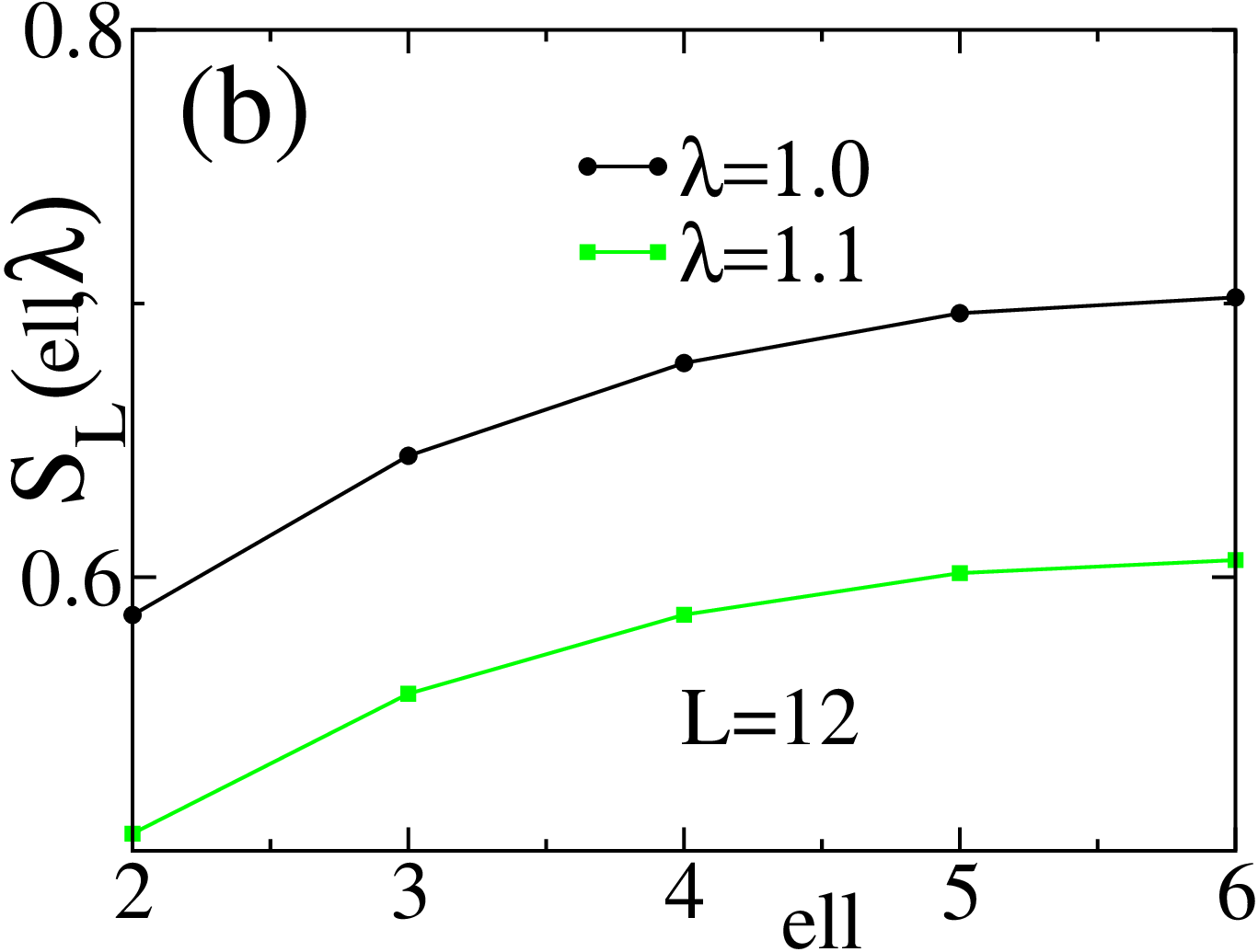}
\par\end{centering}

\begin{centering}
\includegraphics[scale=0.18]{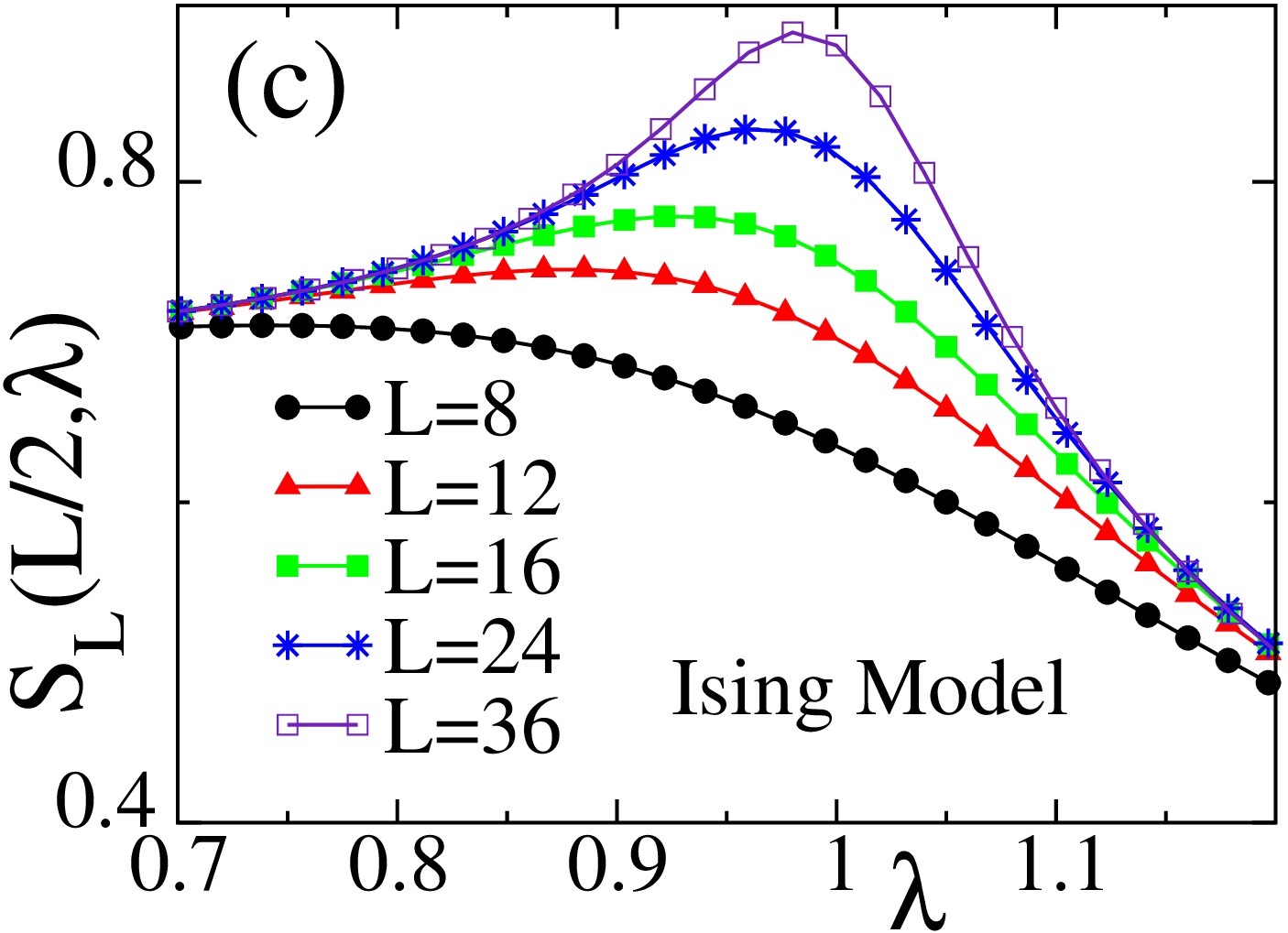}\includegraphics[scale=0.19]{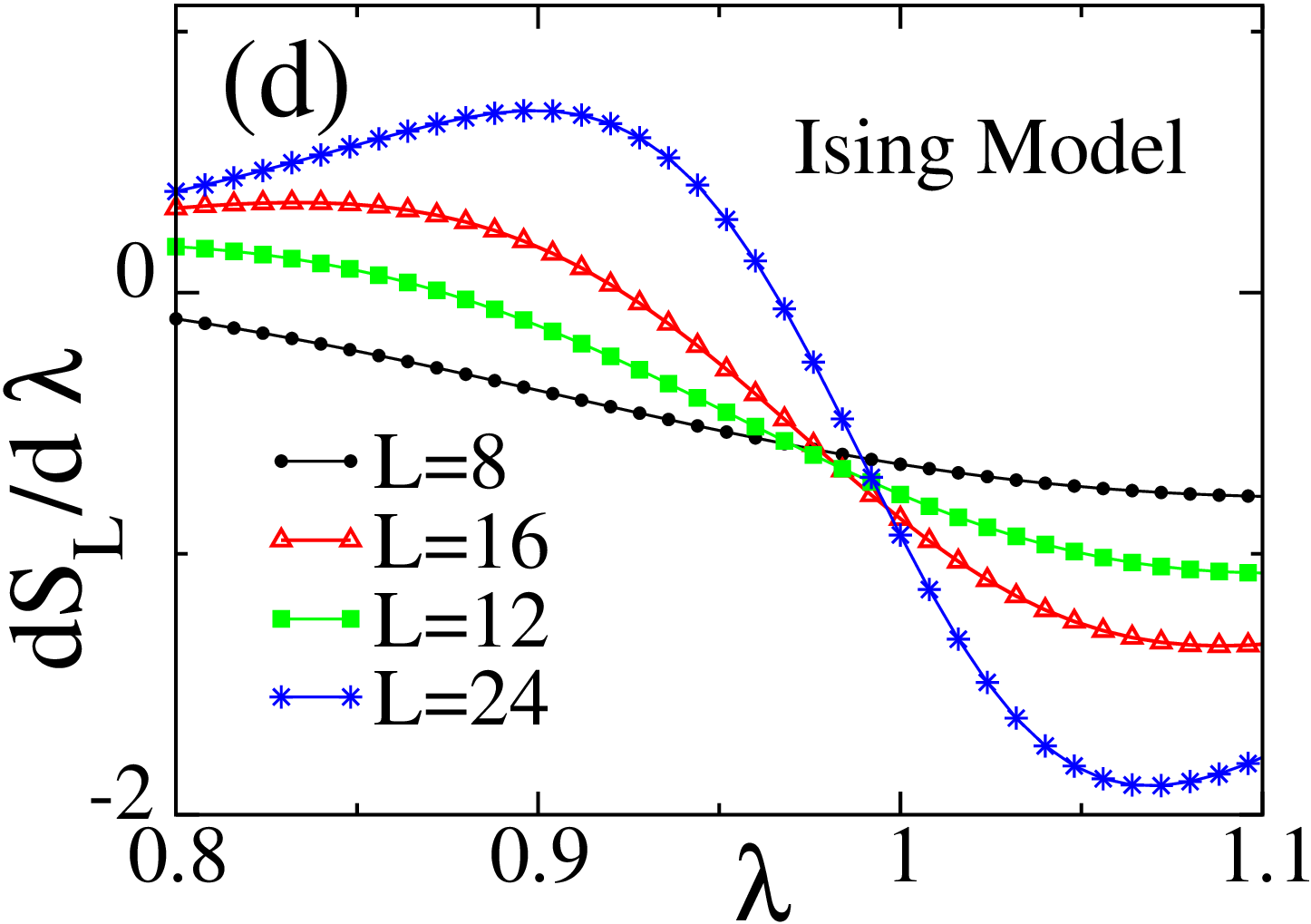}
\par\end{centering}

\begin{centering}
\includegraphics[scale=0.19]{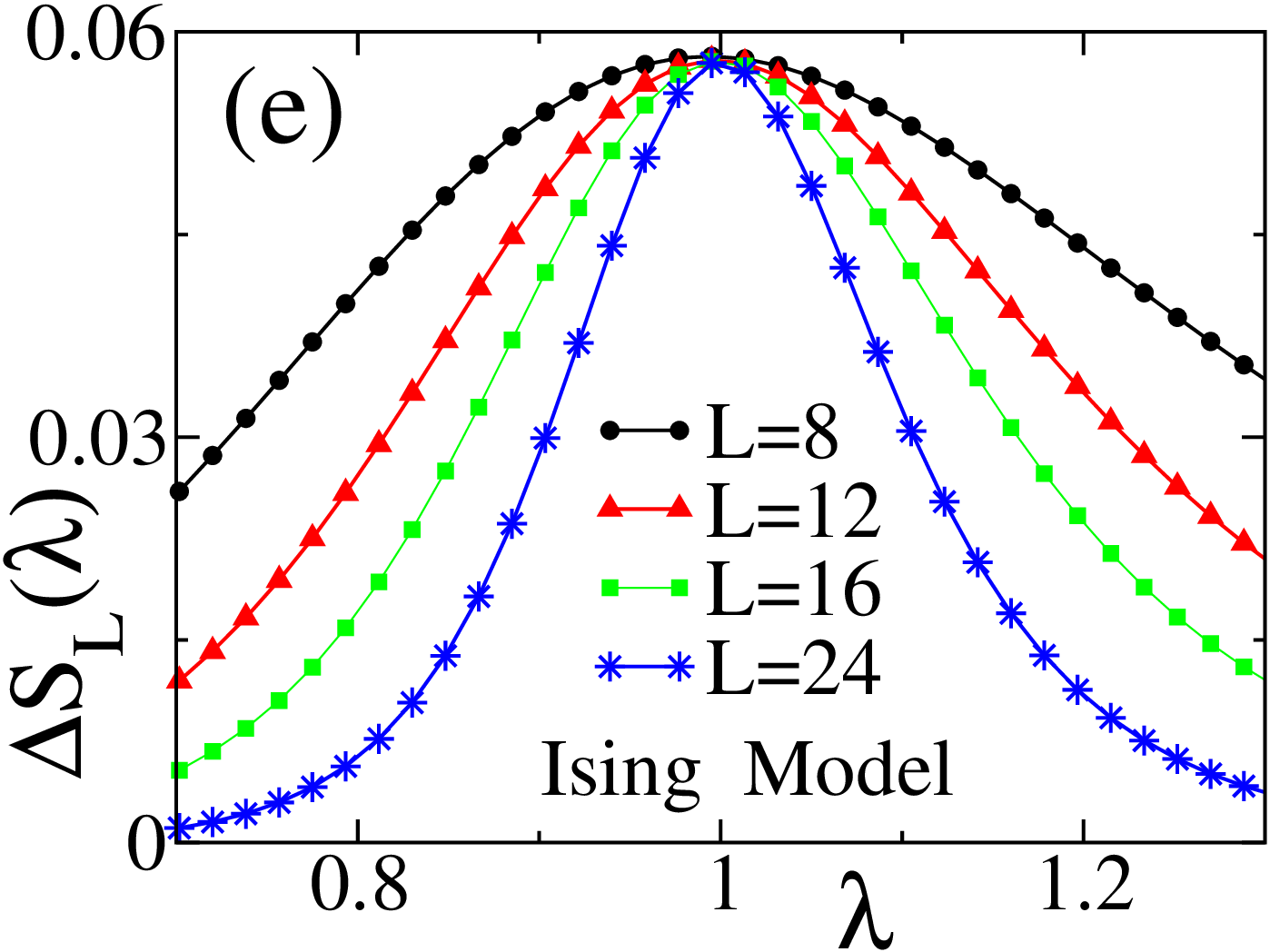} 
\par\end{centering}

\caption{\label{fig1} (Color online). (a) The entanglement entropy $S_{L}(\ell,\lambda)$
vs $\ell$ for the Ising model for a system with size $L=64$ and
some values of $\lambda$ (see legend). (b) Same as (a) but for $L=12$.
(c) $S_{L}(L/2,\lambda)$ vs $\lambda$ for several values of $L$.
(d) $dS_{L}(L/2,\lambda)/d\lambda$ vs $\lambda$. (e) $\Delta S_{L}(\lambda)$
vs $\lambda$. }

\end{figure}

Let us show first the difficulty of extracting the localization of
the critical point by directly using the Eq. (\ref{eq:entropy}).
In Fig.~1(a), we present for the lattice size $L=64$ the entanglement
entropy $S_{L}(\ell,\lambda)$ of the Ising model for $\lambda=1.0,1.01$
and $\lambda=1.1$. These results can be obtained from free fermion
technique.\citealp{PhysRevB.64.064412,0305-4470-36-14-101} We also
use these results to test the precision of our density matrix renormalization
group (DMRG) calculation. For the critical coupling $\lambda=\lambda_{c}=1$
the entanglement entropy has a quite good fit with the expected behavior
given in Eq. (\ref{eq:entropy}) with the values $c=0.5003$ and $\beta=0.4781$,
which are close to the exact ones. \cite{igloiising2008}
On the other hand, for a point clearly away from the critical point,
like $\lambda=1.1$, the entropy tends towards a constant as $\ell$
increases, as expected from Eq. (\ref{eq:entropy}). Although we see
a clear distinct behavior of the entropy for such anisotropies, for
anisotropies closer to the critical point it is quite difficult to
distinguish such behavior even for relatively large lattices. For
example, as we can see in Fig.~1(a), it is difficult to discern if
we are at the critical point or not at $\lambda=1.01$ for a chain
with lattice size $L=64$. In fact, for this coupling, we also obtain
a nice fit with the predicted critical behavior given in 
Eq. (\ref{eq:entropy}).
The fit to our numerical data gives us\textbf{ $c=0.489$ }and $\beta=0.472$,
which are close to the values at the critical point. The reason, for
not seeing a saturation, as we increase $\ell$, for $\lambda=1.01$,
is that at this coupling and for such lattice size the correlation
length $\zeta_{L}(\lambda)=1/G_{L}(\lambda)>L$. Actually, it is possible
to distinguish the critical and the non critical behavior only when
one consider lattice sizes $L>>\zeta$. This means that for relative
small lattices ($L\sim20$), differently from the CM, it is not possible
to obtain reasonable estimates for the critical coupling constants
only from the behavior of $S_{L}(\ell,\lambda)$. Indeed, as shown
in Fig.~1(b), for the lattice size $L=12$, it is even difficult
to distinguish the behavior at the anisotropies $\lambda=1$ and $\lambda=1.1$.
Certainly, the distinct behavior of $S_{L}(\ell,\lambda)$ as a function
of $\ell$, should be more evident at $\ell=L/2$. So, let us check
if the signature of the critical point is more evident for $\ell=L/2$.
In Fig.~1(c), we present $S_{L}(L/2,\lambda)$ as a function of $\lambda$
for lattice sizes $L=8,12,16,24$ and $36$. In this figure only for
$L\geq24$ we start to see a clear change of behavior around the expected
critical point. We observed that the maximum occurs at $\lambda\approx0.97$.
This means that the simple use of $S_{L}(L/2,\lambda)$ will give
an acceptable estimate only for quite larger lattices, as compared
with those necessary to obtain reasonable estimates through the use
of the CM. The fact that we need to consider large systems to observe
in $S_{L}(L/2,\lambda)$ the signature of the critical point is not
particular of the Ising model. This behavior have been observed also
in other models \citep{PhysRevB.83.195105,PhysRevB.77.214418,PhysRevB.83.113104}
{[}see also Fig. 2(a){]}. We should also mention that we did not observe
a clear signature of the quantum critical point, for small lattice
sizes, in the derivative of the entanglement entropy $\frac{dS(L/2,\lambda)}{d\lambda}$
{[}see Fig. 1(d){]}.

Now, we are going to show that $\Delta S_{L}(\lambda)$, defined in
Eq. (\ref{eq:diff}), already presents the signature of the phase
transition for small system sizes. This fact is highly desirable,
because it will not need a huge numerical effort to detect the localization
of the phase transition. In Fig.~1(e), we depict the difference $\Delta S_{L}(\lambda)$
as a function of $\lambda$, for the same lattice sizes presented
in Fig.~1(c). We notice clearly two interesting features: (i) the
maximum of $\Delta S_{L}(\lambda)$ appears close to the true critical
point $\lambda_{c}=1$, and more important (ii) the signature of the
phase transition is observed already for small lattices, as happens
in the CM (see Table I).

\begin{table}
\begin{tabular}{ccc}
\multicolumn{1}{c}{$L$} & $\lambda_{c}^{L}$  & $c^{L}$\tabularnewline
\hline
\hline 
8  & 0.99307  & 0.50374\tabularnewline
 & (1.00197) & \vspace*{.2cm}\tabularnewline
12  & 0.99813  & 0.50137\tabularnewline
 & (1.00049) & \vspace*{.2cm}\tabularnewline
16  & 0.99939  & 0.50073\tabularnewline
 & (1.00014) & \vspace*{.2cm}\tabularnewline
24  & 0.99984  & 0.50031\tabularnewline
 & (1.00005) & \vspace*{.2cm}\tabularnewline
36  & 0.99996  & 0.50014\tabularnewline
 & (1.00001) & \vspace*{.2cm}\tabularnewline
48  & 0.99997  & 0.50008\tabularnewline
 & (1.00001) & \vspace*{.2cm}\tabularnewline
\hline 
exact  & 1.0  & 0.5\tabularnewline
\end{tabular}

\caption{The finite-size estimates of $\lambda_{c}^{L}$ and $c^{L}$ for the
quantum Ising model acquired from the MVEED method and CM. The results
in parentheses are from the CM. The exact values of these quantities
are also shown. }

\end{table}

In Table I, we present the finite-size estimates of $\lambda_{c}^{L}$
obtained from the maximum value of $\Delta S_{L}(\lambda)$.  We also
show in this table the finite-size estimate of the central charge
$c^{L}=6\Delta S_{L}(\lambda_{c}^{L})/\ln(2)$. As we can note, it
is grateful to see that accurate estimates, which are around $0.01\%$
of the exact value ($\lambda_{c}=1$, $c=1/2$), are already obtained
for the small lattice size $L=8$. We also show in Table I, for comparison,
the estimated values acquired from the standard CM {[}see Eq. (\ref{eq:1}){]}.
We learn from this application that the results obtained with the
MVEED method are as precise as the standard CM. Differently from the
CM, our method has the advantage to give us, as a bonus, good estimates
of the conformal anomaly $c$ at the same time we got the estimate
$\lambda_{c}^{L}$. In order to see that the success of this method
in not a particular case of the exactly integrable quantum Ising chain,
in the next section we present results for a non integrable quantum
chain.

\section{Blume-Capel Quantum chain - critical point determination}

The Blume-Capel quantum chain is obtained by the time-continuum limit
of the Blume-Capel model in two dimensions. It describes the dynamics
of spin-1 localized particles, with Hamiltonian given by

\begin{equation}
H_{BC}=-\sum_{j}\left(s_{j}^{z}s_{j+1}^{z}-\delta(s_{j}^{z})^{2}-\gamma s_{j}^{x}\right),\label{eq:BC}\end{equation}
where $s^{x}$ and $s^{z}$ are the spin-1 operators. This Hamiltonian,
differently from the quantum Ising chain, has two coupling constants
($\delta$ and $\gamma$) and is not exactly integrable. The phase
diagram in the plane $\delta-\gamma$ is known from earlier numerical
studies based in the CM (see Fig. 1 of Ref.  \onlinecite{PhysRevB.32.7469}).
For values $\gamma>\gamma_{\mbox{\scriptsize tr}}$ the Hamiltonian
has a quantum critical line $\delta_{c}(\gamma)$ governed by a CFT
in the same universality class of the quantum Ising chain ($c=1/2$).
At $\gamma_{\mbox{\scriptsize tr}}$ the model has a quantum tricritical
point at $\delta_{\mbox{\scriptsize tr}}$ in the universality class
of the tricritical Ising model, having central charge $c=7/10$. For
$\gamma<\gamma_{\mbox{\scriptsize tr}}$ there is a line $\delta_{g}(\gamma)$
of first-order phase transitions (gapped). The numerical estimate
of the tricritical point ($\gamma_{\mbox{\scriptsize tr}},\delta_{\mbox{\scriptsize tr}}$)
was acquired by a generalization of the CM. It was obtained either
by the crossing of the two lowest mass gaps of two lattice sizes,
or by the crossing of the first mass gap of three lattice sizes. These
previous results were obtained for small lattice sizes ($L\leq9$)
and the mass gaps were calculated by using the Lanczos method.\citealp{dagottorev}
In this work, we also used Lanczos method together with the CM to
extend the previous finite-size estimates of $(\gamma_{c},\delta_{c})$
up to lattice size $L=15$. Some of these estimates are presented
in Table II.

Now, let us calculate the critical line of the model by using the
MVEED {[}Eq. (\ref{eq:diff2}){]}.  Although it is also possible to
calculate the entanglement entropy with the Lanczos method, here,
we used the DMRG technique\citealp{white} to do this task. In the
DMRG the entanglement entropy is easily obtained since the reduced
density matrix is naturally calculated in all sweeps of the DMRG.
In our DMRG, we kept up to $m=600$ states per block in the final
sweep. We have done $\sim4-6$ sweeps, and the discarded weight was
typically $10^{-12}-10^{-15}$ at that final sweep.

\begin{figure}[!t]
\begin{centering}
\psfrag{$\ell$}{$\ell$}
\includegraphics[scale=0.18]{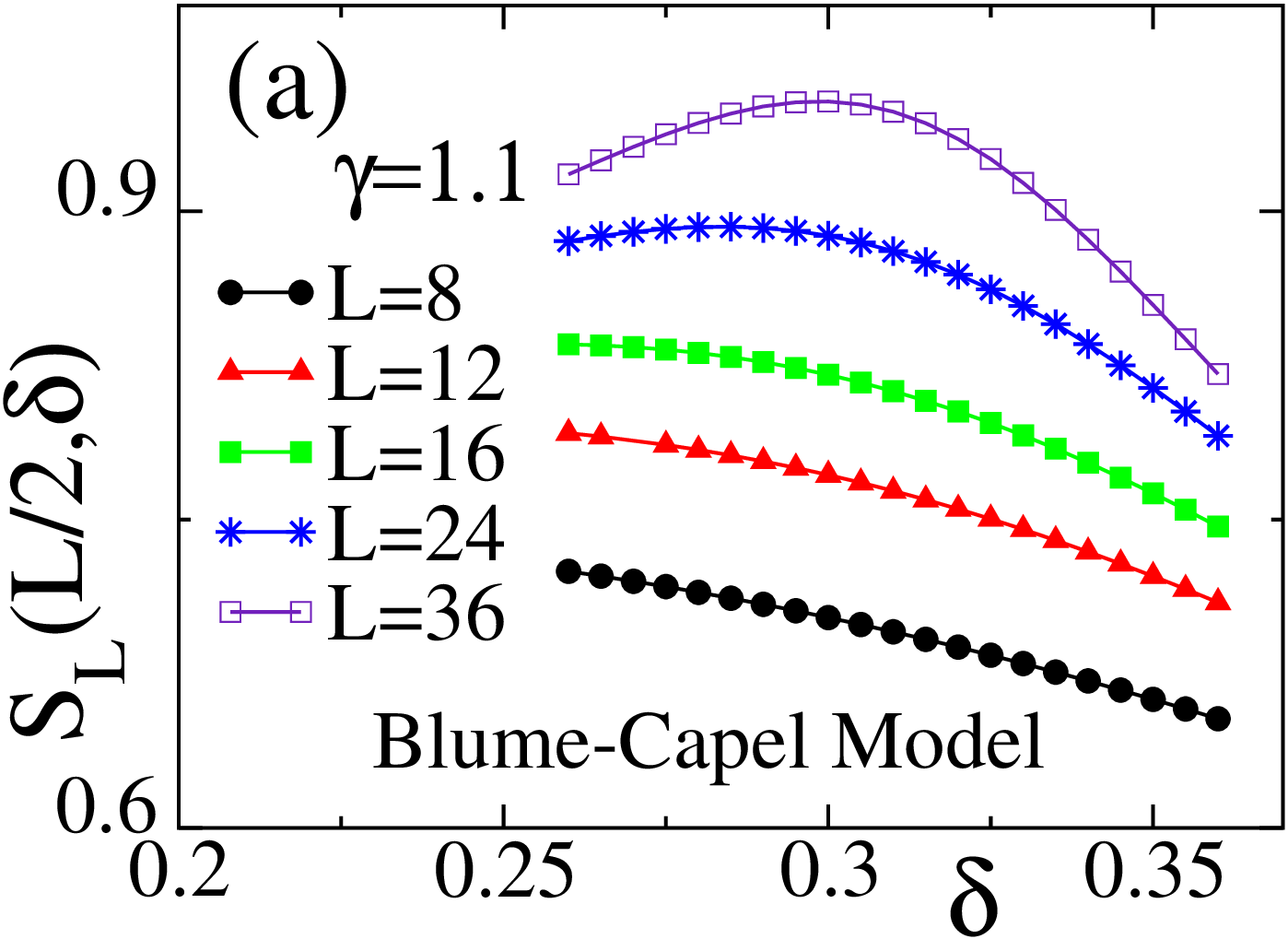}\includegraphics[scale=0.18]{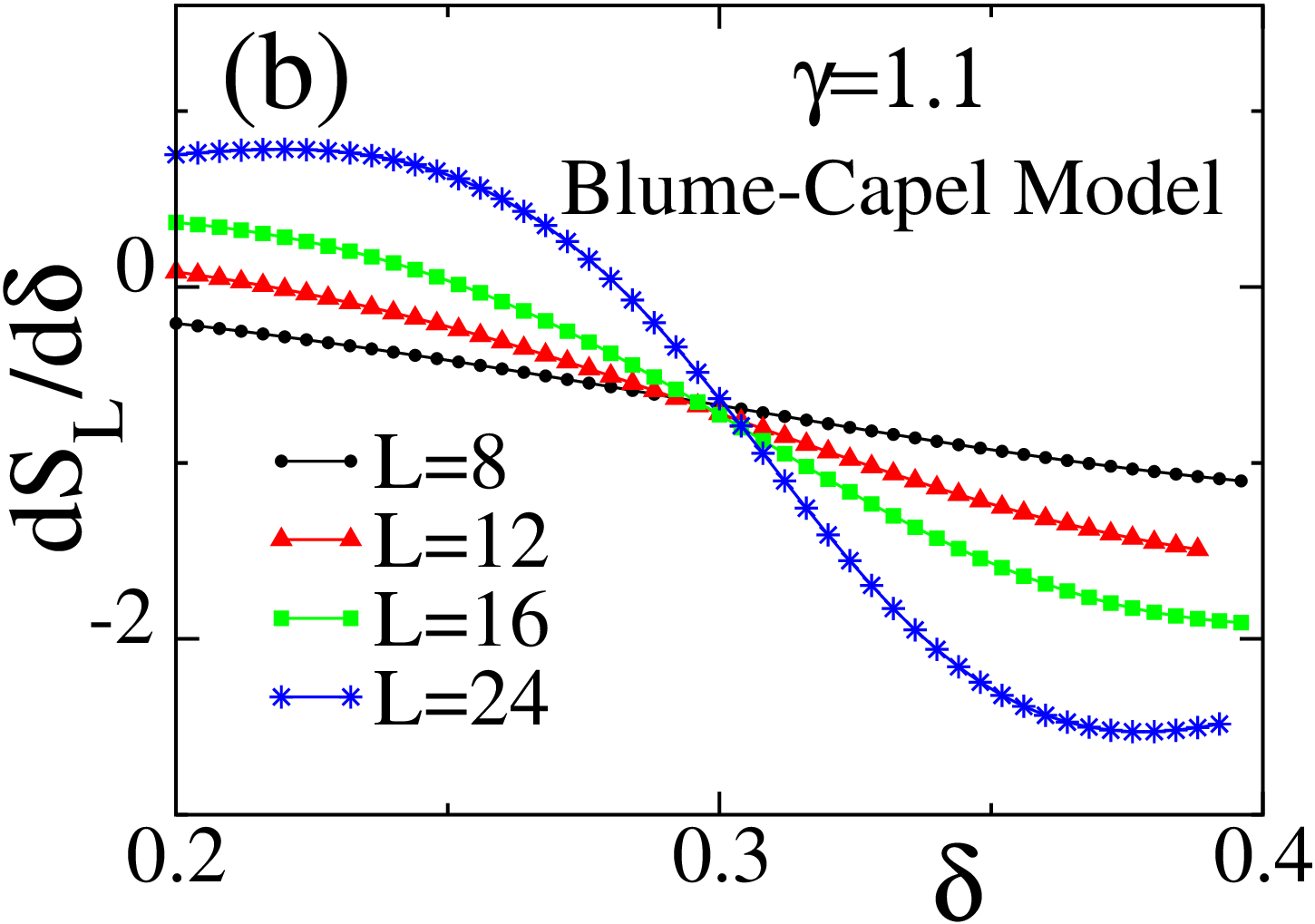}
\par\end{centering}

\begin{centering}
\includegraphics[scale=0.18]{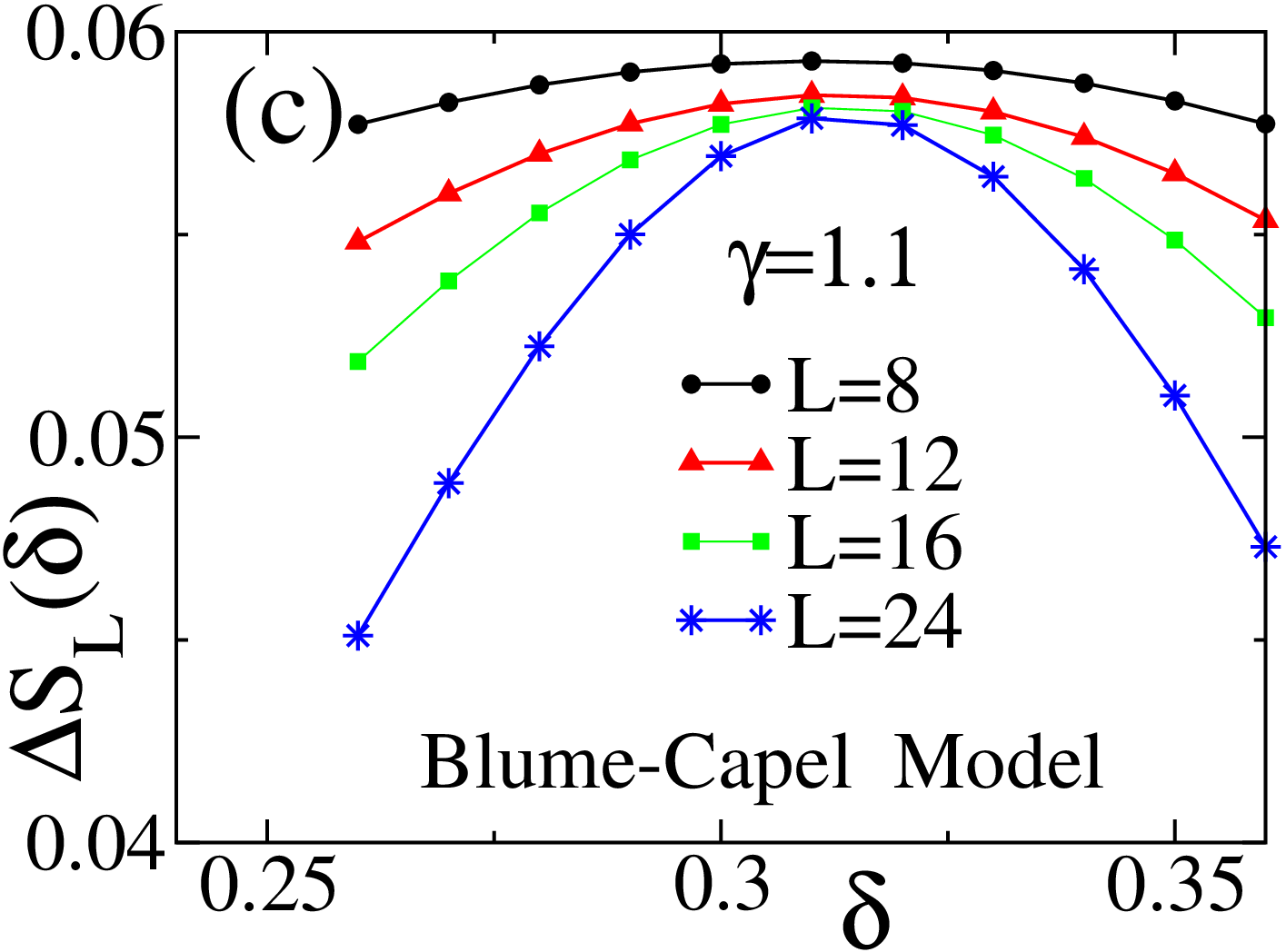}\includegraphics[scale=0.18]{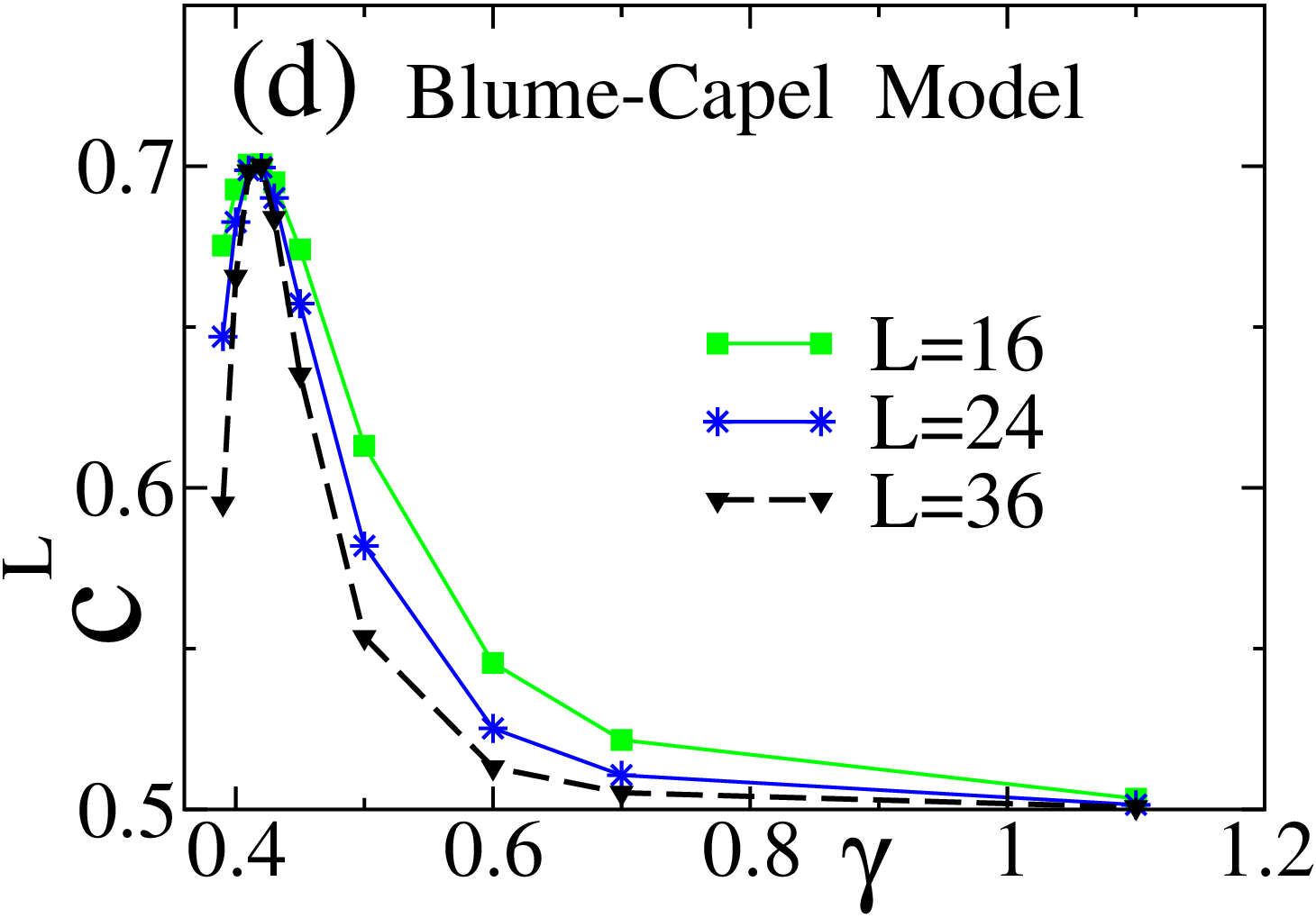}
\par\end{centering}

\begin{centering}
\includegraphics[scale=0.18]{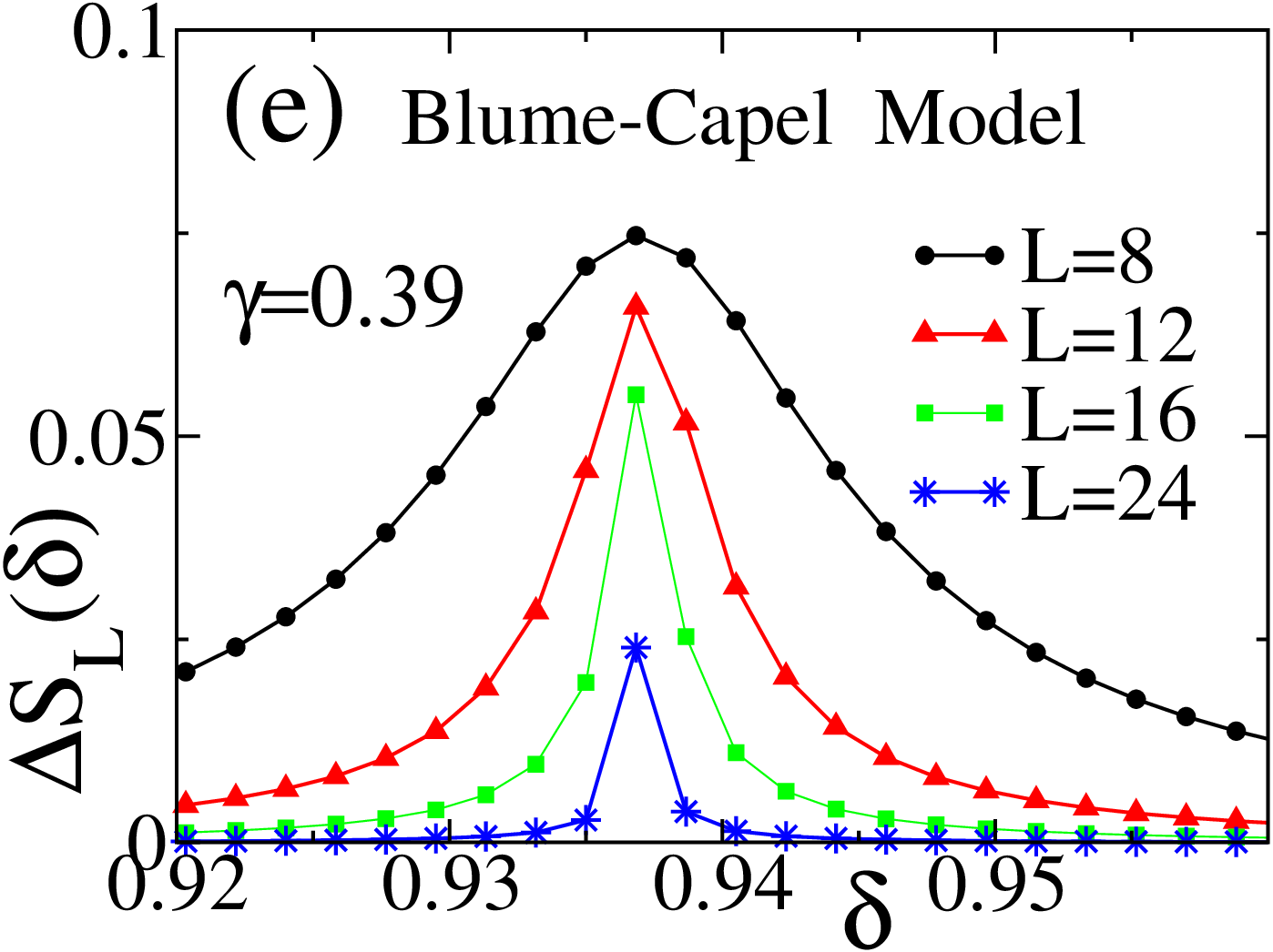}
\par\end{centering}

\caption{\label{fig2} (Color online). (a) The entanglement entropy $S_{L}(\ell=L/2,\delta)$
vs $\delta$ for the spin-1 Blume-Capel model with $\gamma=1.1$ and
for several values of $L$. (b) $dS_{L}(L/2,\delta)/d\delta$ vs
$\delta$. (c) $\Delta S_{L}(\delta)$ vs $\delta$ for the spin-1
Blume-Capel model with $\gamma=1.1$. (d) Finite-size estimates of
$c^{L}$ as function of $\gamma$ for three values of $L$. (e) Same
as (b) but for $\gamma=0.39$.}

\end{figure}

Like in the case of the quantum Ising chain, let us consider first
the behavior of the entanglement entropy for the subsystem with size
$\ell=L/2$, i. e., $S_{L}(L/2,\gamma,\delta)$. We consider initially
a value of $\gamma$ where a second-order phase transition in the
Ising universality is expected. This is the case for $\gamma=1.1$,
where a critical point happens for $\delta_{c}\simeq0.3135$ (see
Table II ).

In Fig.~2(a), we show the entropy $S_{L}(L/2,\gamma=1.1,\delta)$
as a function of $\delta$. Similarly as in the application for the
quantum Ising chain, this entropy does not present any trace of a
quantum phase transition \emph{for small system sizes}. We start to
observe the signature of the phase transition only for relative large
lattice sizes ($L\geq24$). Similar results is also seen in the derivative
$\frac{d S_{L}(\gamma,\delta)}{d\delta}$, as depicted in Fig.
2(b) for $\gamma=1.1$. On the other hand, if we now use our estimator
$\Delta S_{L}(\gamma,\delta)$, as shown in Fig.~2(c), we already
see a clear appearance of a peak for the small lattice size $L=8$.
The finite-size estimates $\delta_{L}^{c}$, for three values of $\gamma$,
obtained from the MVEED method are presented in Table II. It is also
given in this table, for comparison, the  values of $\delta_{L-2,L}^{c}$
obtained through the standard CM {[}Eq. (\ref{eq:1}){]}. We clearly
see in the table that the results acquired from the MVEED method converged
already to almost five digits for small lattice sizes like $L=24$.

%
\begin{table}
\begin{tabular}{cccc}
\multicolumn{3}{c}{\hspace*{1cm}$\delta_{L}^{c}$ } & \tabularnewline
\hline 
 & $\gamma=1.1$  & $\gamma=0.41$  & $\gamma=0.39$\tabularnewline
\hline 
8  & 0.31072  & 0.91260  & 0.92121\tabularnewline
 & (0.30997)  & (0.91317)  & (0.92240)\vspace*{.2cm}\tabularnewline
12  & 0.31279  & 0.91271  & 0.92132\tabularnewline
 & (0.31254)  & (0.91292)  & (0.92190)\vspace*{.2cm}\tabularnewline
16  & 0.31321  & 0.91273  & 0.92133\tabularnewline
 & (0.31315)  & (0.91286)  & (0.92171)\vspace*{.2cm}\tabularnewline
24  & 0.31351  & 0.91274  & 0.92134\tabularnewline
  & (0.31346)  & (0.91281)  & (0.92155)\vspace*{.2cm}\tabularnewline
36  & 0.31355  & 0.91274  & 0.92135\tabularnewline
 & (0.31354)  & (0.91278)  & (0.91245)\vspace*{.2cm}\tabularnewline
48  & 0.31357  & 0.91275  & 0.92136\tabularnewline
 & (0.31356)  & (0.91277)  & (0.92140)\vspace*{.2cm}\tabularnewline
\hline 
\end{tabular}

\caption{The finite-size estimates of $\delta^{c}_{L}$ for the Blume-Capel model
for three values of $\gamma$ obtained from the MVEED method and the
CM. The results in parentheses are from the CM. }

\end{table}

Note that the results of Table II tell us that the estimated results
obtained from the MVEED method are as good, if not better, than those
coming from the CM. This table also includes the extrapolated results
acquired by using the VBS method. \citep{VBS}

As we already mentioned, at the same time we get the finite-sizes
estimates of $\delta_{L}^{c}$, we also obtain the finite-size estimates
of the central charge $c^{L}$. In Fig.~2(d), we depict these estimates
for the lattice sizes $L=16,24$ and $36$ for several values of 
$\gamma(\delta^{c})$.
We clearly see, in this figure, that the estimates of $c^{L}$ tend
to $c=1/2$ for $\gamma>0.6$ and, for $0.4<\gamma<0.6$, exhibit
a quite large change with a peak in a value close to $c^{L}=7/10$,
which is the expected value of the central charge for the tricritical
point of the BCM.\citep{xavierbc}

It is interesting to note that like the CM, the method we propose
in this paper also gives us estimates of critical quantum points,
in regions where they do not exist (gapped regions), like the ones
presented in Fig.~2(d) for $\gamma\lesssim0.4$. This happens in
both methods due to the fact that correlation length is large than
the system size considered, i. e., $\zeta=1/G_{L}>L$ in this region.
However, differently from the CM the MVEED method, at least in this
application, allows us to distinguish these fake results. In the case
of a true critical point the maximum of $\Delta S_{L}$, as $L$ increases,
should saturate on a finite value, while in the non critical case
it should decrease to zero. In Fig.~2(e), we show $\Delta S_{L}$
as a function of $\delta$ for $\gamma=0.39$. For this coupling we
expect a first order phase transition, with a small gap. It is clear
that for quite small lattices like $L=8$ and $L=12$ we observe the
peak of $\Delta S_{L}$ tending towards zero.

Our estimate for the tricritical point ($\gamma_{\mbox{\scriptsize tr}},\delta_{\mbox{\scriptsize tr}}$)
comes from the maximum value of $\Delta S_{L}(\gamma,\delta)$ (in
the $\gamma$-$\delta$ plane). In Table III, we present the estimated
values of the tricritical couplings obtained by this procedure. We
also show in this table the finite-size estimates of the central charge
$c^{L}$ obtained with the MVEED method. Notice the quite good convergence
towards the predicted value $c=7/10$.\textbf{ }One finite-size estimate
of the tricritical point obtained from the CM is also presented. 

The evaluation of those tricritical points, by using the CM, is much
more difficult since in this case we should find the crossing of two
gaps in a pair of lattice sizes or use the crossing of the first gap
in three lattice sizes. This means that to obtain a sequence with
a given number of estimates we need larger lattice sizes in the CM.
The available finite-size estimates derived from the CM (up to $L=8$)
are of the same order of precision as those obtained by the MVEED
method by using the same lattice sizes. However, in the MVEED method
we get a larger sequence to extrapolated, since in this case there
is one estimate for each lattice size. For this reason, in principle,
we should obtain more precise extrapolated values.

\begin{table}
\begin{tabular}{cccc}
L  & $\gamma^{tric}$  & $\delta^{tric}$  & $c^{L}$\tabularnewline
\hline 
8  & 0.41047  & 0.91240  & 0.7065\vspace*{.2cm}\tabularnewline
12  & 0.41405  & 0.91091  & 0.7025\vspace*{.2cm}\tabularnewline
16  & 0.41496  & 0.91052  & 0.7014\vspace*{.2cm}\tabularnewline
24  & 0.41551  & 0.91028  & 0.7008\vspace*{.2cm}\tabularnewline
36  & 0.41560  & 0.91024  & 0.7003\vspace*{.2cm}\tabularnewline
Ref. {[}\onlinecite{PhysRevB.32.7469}{]}  & 0.41563  & 0.91024  & ----\tabularnewline
\end{tabular}

\caption{The finite-size estimates of $\gamma^{tric}$ and $\delta^{tric}$ , and
$c^{L}$ for the tricritical point of the Blume-Capel model obtained
with the MVEED. The finite-size estimate of the tricritical couplings
taken from the Ref. \onlinecite{PhysRevB.32.7469} for sizes $L=8,9$
is also presented. }

\end{table}

\section{Conclusions}

We have introduced a practical and simple method, called maximum value
of the entanglement entropy difference (MVEED), to estimate the quantum
critical points of quantum chains. The MVEED method explores the distinct
behaviors of the entanglement entropy at critical (gapless) and non
critical (gapped) couplings. We made two applications of this method.
We calculated the critical points for the quantum Ising chain and
for the quantum Hamiltonian of the Blume-Capel model. We compared,
in both models, the obtained results with those coming from the standard
crossing method (CM). Our results show clearly that the MVEED method
gives us accurate estimates for small lattice sizes, similar as those
obtained with the standard CM. However, the MVEED method has some
advantages as compared with the CM:

(i) \textit{Simplicity of calculation}. In order to obtain a finite-size
estimate for the critical coupling constants with the MVEED method
we \emph{only} need the eigenfunction of the ground state, in contrast
with the CM where it is necessary to calculate the mass gaps (two
different energies) for two distinct lattice sizes. For this reason
the CM demands more computational effort, as compared with the MVEED
method. 

(ii) \textit{Identification of the universality class}. Differently
from the CM, at the same time the critical point is obtained the central
charge $c$, that identify the universality class of the critical
behavior, is also calculated. In the usual finite-size scaling where
we use the CM, the central charge can also be estimated by using the
consequences of conformal invariance in the eigenspectra of the finite
quantum chain.\citep{anomaly2,anomaly1} We need, in this case, to
evaluate additional mass gaps, in order to estimate the sound velocity.

(iii) \textit{Effectiveness for the localization of tricritical points.}
Due to the property (ii) the method is quite effective to locate tricritical
points since the universality class of critical behavior changes at
the tricritical points. This was illustrated in the example of the
BCM (section 4). In the usual CM the numerical effort is higher, because
the finite-size estimates are obtained by using the two lowest mass
gaps of two lattice sizes or the lowest mass gaps of three distinct
lattice sizes.

(iv) \textit{Sensibility to detect the first-order phase transitions.}
Gapped quantum chains with small gaps ($\zeta_{L}=1/G_{L}>L$), produce
crossing points in the usual CM, but also a maximum in the proposed
estimator $\Delta S_{L}$. In the case of the CM this can be decided
only by further analysis of the finite-size dependence of the mass
gaps, and usually it is necessary large lattices. Surprisingly in
the application we did for the BCM, our proposed method shows, already
for quite small lattice sizes, a distinct behavior for the gapped
and non-gapped systems. Even for non-critical points that are close
to the tricritical point (small mass gaps) the estimator, instead
of saturating in a non-zero value, goes to zero, as we should expect
in a gapped system. This is a surprise because, similarly as the CM,
the proposed method is also based on the variations of the correlation
length, which are quite small. However, distinct form the CM, our
method shows a large sensitivity already in quite small lattice sizes.
Future applications of the proposed method will confirm if this is
a general feature of the method or an accident of the present application. 

It is interesting to stress that differently from several existing
estimators, based on the entanglement properties, the MVEED method
\emph{was tested in small lattice sizes} producing surprisingly good
results. For the above reasons, we believe the use of this method
may become popular in future applications.

In order to conclude, we mention that a generalization of the presented
method, based on the distinct behavior of the entanglement entropy,
to locate quantum critical points, can also be introduced for higher
dimensions at least for fermionic systems with a finite Fermi surface.
In this case, a natural estimator would be $\Delta S_{L}/L^{d-1}.$ 
\begin{acknowledgments}
The authors thank M. Sarandy and G. Rigolin for useful discussions.
This research was supported by the Brazilian agencies FAPEMIG, FAPESP,
and CNPq. 
\end{acknowledgments}
\bibliographystyle{apsrev4-1}

\begin{thebibliography}{10}
\bibitem{commgap}%
  \BibitemOpen
   \bibinfo {year}  \bibinfo {note}{Actually, for general quantum
  critical points we should have the crossing of $L^zG(L)$, where $z$ is the
  dynamical critical exponent. However, most of the quantum critical chains are
  conformally invariant, and in this case $z=1$.}%
  \bibAnnoteFile{Stop}{commgap}%
\bibitem{commx} The estimate obtained by considering 
$ \Delta S_{LK}$ ( $\Delta S_{L}$) with $L=16$ for the Ising 
quantum chain is $\lambda_c=0.98279$ ($\lambda_c=0.99939$) and for the
Blume Capel Model with $\gamma=1.1$ is $\delta_c=0.29809$ ($\delta_c=0.31321$).
\end{thebibliography}
%

\end{document}